\documentclass[smallextended]{svjour3}       
\smartqed  
\usepackage{graphicx}
\begin{document}

\title{Presentism meets black holes
}


\author{Gustavo E. Romero \and Daniela P\'erez}

\authorrunning{Gustavo E. Romero \and Daniela P\'erez} 
\institute{Gustavo E. Romero \and Daniela P\'erez
						\at Instituto Argentino de Radioastronom{\'\i}a, C.C.5, (1984)\\
              Villa Elisa, Bs. As., Argentina \\
              \email{romero@iar-conicet.gov.ar,\\
 							danielaperez@iar-conicet.gov.ar} 
							\and
							Gustavo E. Romero 
							\at Facultad de Ciencias Astron\'omicas y Geof{\'\i}sicas, UNLP, Paseo del Bosque s/n\\
							CP (1900), La Plata, Bs. As., Argentina}

\date{Received: date / Accepted: date}

\maketitle

\begin{abstract}
Presentism is, roughly, the metaphysical doctrine that maintains that whatever exists, exists in the present. The compatibility of presentism with the theories of special and general relativity was much debated in recent years. It has been argued that at least some versions of presentism are consistent with time-orientable models of general relativity. In this paper we confront the thesis of presentism with relativistic physics, in the strong gravitational limit where black holes are formed. We conclude that the presentist position is at odds with the existence of black holes and other compact objects in the universe. A revision of the thesis is necessary, if it is intended to be consistent with the current scientific view of the universe.

\end{abstract}

\section{Introduction}

Presentism is a metaphysical thesis about what there is. It can be expressed as (e.g. Crisp 2003):
\begin{quotation}
{\it Presentism}. It is always the case that, for every $x$, $x$ is present. 
\end{quotation}
The quantification is unrestricted, it ranges over all existents. In order to make this definition meaningful, the presentist must provide a specification of the term `present'. Crisp, in the cited paper, offers the following definition:
\begin{quotation}
{\it Present}. The mereological sum of all objects with null temporal distance. 
\end{quotation} 
The notion of temporal distance is defined loosely, but in such a way that it accords with common sense and the physical time interval between two events. From these definitions it follows that the present is a thing, not a concept. The present is the ontological aggregation of all present things. Hence, to say that `$x$ is present', actually means ``$x$ is part of the present".   

The opposite thesis of presentism is eternalism, also called four-dimensionalism. Eternalists believe in the existence of past and future objects. The temporal distance between these objects is non-zero. The name four-dimensionalism comes form the fact that in the eternalist view, objects are extended through time, and then they have a 4-dimensional volume, with 3 spatial dimensions and 1 time dimension. There are different versions of eternalism. The reader is referred to Rea (2003) and references therein for a discussion of eternalism.

Several philosophers have defended presentism from various critiques, notably Chisholm (1990), Bigelow (1996), Zimmerman (1996, 2011), Merricks (1999), Hinchliff (2000), Crisp (2007), and Markosian (2004)\footnote{See also the recent review by Mozersky (2011).}. Some of these authors have considered criticisms arisen from the alleged incompatibility of presentism with relativity theory. In what follows we review the main objections posed to presentism in the framework of both relativities, special and general, and then we move on to consider a new type of argument based on the existence of compact objects in the universe. We  show that the presentist cannot accept the standard physical understanding of these objects, without introducing changes in his ontological views or rejecting current astrophysics\footnote{ We note that some authors such as Savitt (2006), Dorato (2006), Dolev (2006), and more recently Norton (2013) have argued that the dispute between presentism and eternalism is not a genuine one. We are not concerned with this dispute, but with the consistency of presentism with general relativity.}.

\section{Presentism and special relativity}

Special relativity is the theory of moving bodies formulated by Albert Einstein in 1905 (Einstein 1905). It postulates the 
Lorentz-invariance of all physical law statements that hold in a special type of reference systems, called {\it inertial frames}. Hence the `restricted' or `special' character of the theory. The equations of Maxwell electrodynamics are Lorentz-invariant, but those of classical mechanics are not. When classical mechanics is revised to accommodate invariance under Lorentz transformations between inertial reference frames, several modifications appear. The most notorious is the impossibility of defining an absolute simultaneity relation among events. The simultaneity relation results to be frame-dependent. Then, some events can be future events in some reference system, and present or past in others. Since what there is cannot depend on the reference frame adopted for the description of nature, it is concluded  that past, present, and future events exist. Then, presentism is false.

This argument has been formulated with different levels of sophistication by philosophers such as Smart (1963), Rietdijk (1966), and Putnam (1967). In particular, both Rietdijk and Putnam argued independently that special relativity implies determinism. This position was revisited by Maxwell (1985) in the eighties. He maintained that probabilism\footnote{Probabilism is the thesis that the universe is such that, at any instant, there is only one past but many alternative futures (Maxwell 1985).} and special relativity were incompatible. Strong objections to Rietdijk, Putnam, and Maxwell can be found in Stein (1968, 1991)\footnote{The controversy between Putman and Stein is reviewed by Saunders (2002).}. In his 1991 paper Stein showed that ``the openness of the future relative to the past for any region of space-time is, given special relativity, logically incompatible with the objective global temporal structure'' (Harrington 2009). In later years Harrington (2008, 2009) defended a theory of local temporality, also called point-present theory, which assuming special relativity allows for the possibility of an open future, i.e not fully determined (see also Ellis (2006) and Ellis and Rothman 2010).

The argument for eternalism from special relativity seems particularly strong in the reformulation of this theory proposed by Hermann Minkowski. Minkowski (1907, 1909) introduced the concept of space-time: the mereological sum of all events. Space-time can be represented by a 4-dimensional manifold endowed with a metric that allows to compute distances between events. These are spatio-temporal distances or {\it intervals}. The differential (arbitrarily small) interval for Minkowski space-time, which is invariant under Lorentz transformations, is:
	\begin{equation}
	ds^{2}=\eta_{\mu\nu}dx^{\mu} dx^{\nu}=(dx^{0})^{2}-(dx^{1})^{2}-(dx^{3})^{2}-(dx^{3})^{2}, 
	\label{eq:eta}
\end{equation}
where the Minkowskian metric tensor $\eta_{\mu\nu}$ is pseudo-Euclidean, with rank 2 and trace $-2$. The coordinates with the same sign are called {\sl spatial} (the usual convention in notation is $x^{1}=x$,  $x^{2}=y$, and $x^{3}=z$) and the coordinate $x^{0}=ct$ is called {\sl time coordinate}. The constant $c$ is introduced to make the physical units uniform, and can be shown to be equal to the speed of light in empty space. 
%
       
\begin{figure}
\centerline{\includegraphics[natwidth=864,natheight=676,width=30pc]{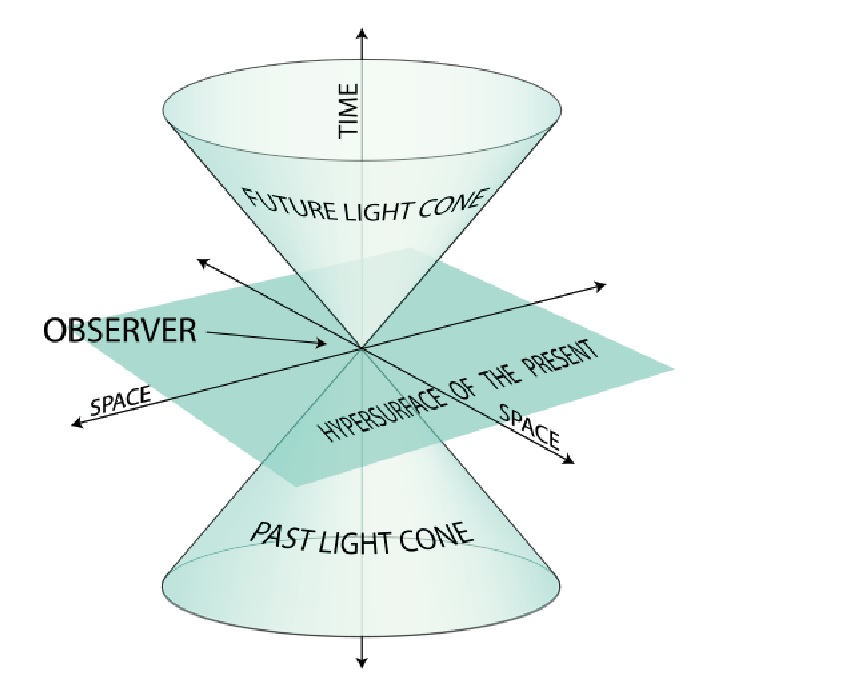}} 
\vspace{2pc}
\caption[]{Light cone. Causal future, past, and the present plane of the central event are shown.}
\label{Fig1}
\end{figure}

Minkowski's interpretation allows to separate space-time, at each point (every point represents an event), in three regions according to $ds^{2}<0$ (space-like region), $ds^{2}=0$ (light-like or null region), and $ds^{2}>0$ (time-like region). Particles that go through the origin can only reach time-like regions. The null surface $ds^{2}=0$ can be inhabited only by particles moving at the speed of light, like photons. Points in the space-like region cannot be reached by material objects from the origin of the {\sl light cone} that can be formed at any space-time point (see Fig. \ref{Fig1}). 



The introduction of the metric allows to define the future and the past {of a given event}. Once this is done, all time-like events can be classified by the relation `earlier than' or `later than'. The selection of a `present' event - or `now' - is entirely conventional. To be present is not an intrinsic property of any event. Rather, it is a secondary, relational property that requires interaction with a conscious being. The extinction of the dinosaurs will always be earlier than the beginning of the Second World War. But the latter was present only to some human beings in some physical state. The present is a property like a scent or a color. It emerges from the interaction of self-conscious individuals with changing things and has not existence independently of them (for more about this point of view, see Gr\"unbaum 1973, Chapter X).

This sounds convincing, but there is a strong ontological assumption: the system of all events is assumed. This provides a simplification that pleases the physicist, but can terrify the philosopher. Presentists, surely, will plainly reject this assumption as outrageous. But do they have an alternative interpretation of special relativity with the same explanatory and predictive power as that of Minkowski's? 

Yes, they have. The alternative was already suggested by Poincar\'e (1902) in the form of the conventionalist thesis. Poincar\'e remarked that it is always possible to save an interpretation based on an Euclidean physical geometry if additional fields are introduced in the theory to adjust all physical quantities (lengths, masses, speeds) to the adopted geometry. In the case of special relativity, this yields Lorentz's theory of moving bodies. Physicist will always prefer the simplest theory, but the presentist can argue that from an ontological point of view, Minkowski's interpretation is far more costly than the presentist one: a 3-dimensional Euclidean and Lorentz-invariant theory. An Euclidean space-time is consistent with presentism since the interval can be written as:

\begin{equation}
	ds^{2}=(cdt)^{2}+dx^{2}+dy^{2}+dz^{2}. 
	\label{eq:eucl}
\end{equation}

Hence, since in a proper system $ds=cd\tau$, and using the invariance of the interval, we get:

\begin{equation}
d\tau^{2}=dt^{2}+\frac{1}{c^{2}}d\vec{x}^{2}=dt^{2}\left(1+\vec{\beta}^{2}\right),
\label{eq:eucl2}
\end{equation}
where 
$$\vec{\beta}=\frac{\vec{v}}{c}.$$
From Eq. (\ref{eq:eucl2}) follows that $d\tau=0$ iff $dt=0$. All events have zero temporal distance in the 3-dimensional Euclidean plane.  A full discussion of such an alternative is presented by Craig (2001). 

An Euclidean geometry can be preserved (to a high price, i.e. introducing distorting fields) in special relativity since both Euclidean and Minkowskian geometries have zero intrinsic curvature. The curvature of a given manifold is given by the Riemann tensor $R^{\alpha\beta\gamma\delta}$, formed with second derivatives of the metric. This tensor is identically zero for any space-time with a constant metric. Since the intrinsic curvature is the same for both Euclidean and Minkowskian geometries, it is always possible to find a {\it global} transformation between them. In the case of more general space-times, represented by manifolds with intrinsic curvature, such transformations can only be local. This leads us to general relativity. 

\section{Presentism and general relativity}

A basic characteristic of Minkowski space-time is that it is `flat': all light cones point in the same direction, i.e. the local direction of the future does not depend on the coefficients of the metric since these are constants. More general space-times, however, are possible. If we want to describe gravity in the framework of space-time, we introduce a pseudo-Riemannian manifold, whose metric can be flexible, i.e. a function of the material properties (mass-energy and momentum) of the physical systems that produce the events represented by space-time points. 

The key to relate space-time to gravitation is the {\sl equivalence principle} introduced by Einstein (1907): at every point $P$ of the manifold that represents space-time, there is a flat tangent surface.


In order to introduce gravitation in a general space-time we define a metric tensor $g_{\mu\nu}$, such that its components can be related to those of a locally Minkowskian space-time defined by $ds^{2}=\eta_{\alpha\beta}d\xi^{\alpha} d\xi^{\beta}$ through a general transformation: 

\begin{equation}
	ds^{2}=\eta_{\alpha\beta}\frac{\partial \xi^{\alpha}}{\partial x^{\mu}}\frac{\partial \xi^{\beta}}{\partial x^{\nu}}dx^{\mu}dx^{\nu}=g_{\mu\nu}dx^{\mu}dx^{\nu}.
\end{equation}

\vspace{0.2cm}
 
In the absence of gravity we can always find a global coordinate system ($\xi^{\alpha}$) for which the metric takes the form given by Eq. (\ref{eq:eta}) everywhere. With gravity, conversely, such a coordinate system can represent space-time only in an infinitesimal neighborhood of a given point. The curvature of space-time means that it is not possible to find coordinates in which $g_{\mu\nu}=\eta_{\mu\nu}$ at {\sl all points on the manifold}. It is always possible, however, to represent the event (point) $P$ in a system such that $g_{\mu\nu}(P)=\eta_{\mu\nu}$ and $(\partial g_{\mu\nu}/\partial x^{\sigma})_{P}=0$.  
 
The key issue to determine the geometric structure of space-time, and hence to specify the effects of gravity, is to find the law that fixes the metric once the source of the gravitational field is given. The source of the gravitational field is represented by the energy-momentum tensor $T_{\mu\nu}$ \index{energy-momentum tensor} that describes the physical properties of material things. This was Einstein's fundamental intuition: the curvature of space-time at any event is related to the energy-momentum content at that event. 





The Einstein field equations can be written in the simple form (Einstein 1915):

\begin{equation}
	G_{\mu\nu} = -\frac{8\pi G}{c^4}\, T_{\mu\nu}, 
	\label{einstein}
\end{equation}
where $G_{\mu\nu}$ is the so-called Einstein's tensor. It contains all the geometric information on space-time. The constants $G$ and $c$ are the gravitational constant and the speed of light in vacuum, respectively.

Einstein's field equations are a set of ten non-linear partial differential equations for the metric coefficients. In Newtonian gravity, otherwise, there is only one gravitational field equation. General relativity involves numerous non-linear differential equations. In this fact lays its complexity, and its richness.

A crucial aspect of general relativity is that 4-dimensional space-time with non-zero curvature is not dispensable anymore. Is this implying that presentism should be abandoned? Not necessarily. 

\begin{figure}
\centerline{\includegraphics[natwidth=739,natheight=775,width=25pc]{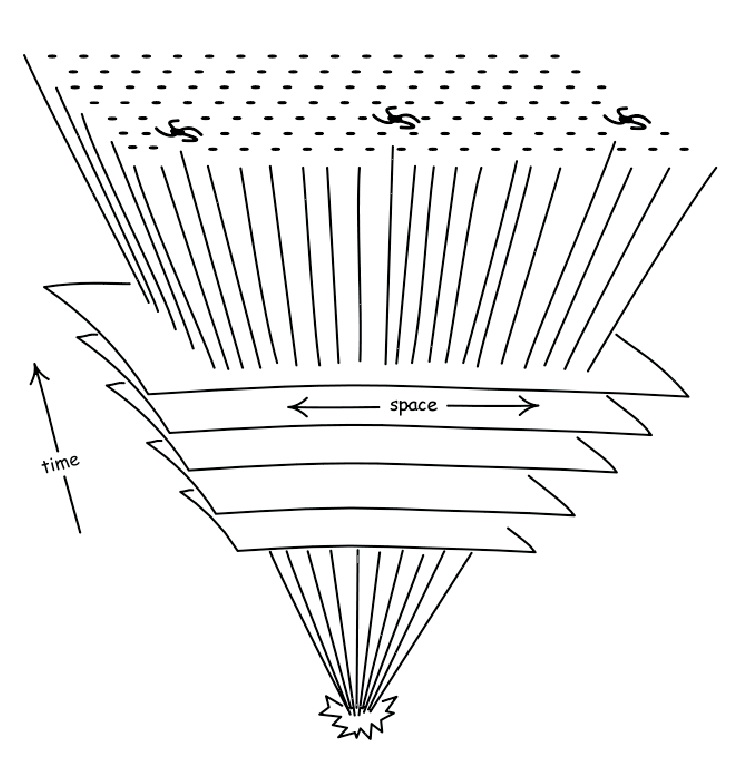}} 
\vspace{2pc}
\caption[]{A `presentist-friendly' space-time: Evolving 3-dimensional space-like surfaces in a space-time with a preferred time-direction.}
\label{Fig2}
\end{figure}

Thomas Crisp (2007) has proposed a ``presentist-friendly'' model of general relativity. He suggests that the world is represented by a 3-dimensional space-like hypersurface that evolves in time. This interpretation requires to introduce a preferred foliation of space-time, and to consider the $3+1$ usual decomposition for the dynamics of space-time in such a way that `the present' is identified with the evolving hypersurface (see Fig. \ref{Fig2}).



In order to formulate such a model of space-time, some global constraints must be imposed: there should be a possible foliation into Cauchy hypersurfaces in order to allow for global time-like continuous vector fields that can be used to introduce a ``global time coordinate''. This is the case, for instance, of the Friedmann-Robertson-Walker-Lema\^itre metric. General conditions for such a metric requires the absence of Cauchy horizons and the fulfillment of the so-called {\it energy conditions} (Hawking and Ellis 1973). 

Although a case can be made for general inhomogeneous metrics and massive violations of the energy conditions on the basis of recent cosmological data (e.g. Pleba\'nski and Krasi\'nski 2006), in what follows we focus on local aspects that should be accommodated in any cosmological model compatible with current astrophysics.

\section{Black holes and the present}\label{BH}

We shall now provide a general definition of a black hole, independently of the coordinate system adopted in the description of space-time, and even of the exact form of the field equations. A space-time is represented by an order pair, formed by a manifold $M$ and a metric $g_{\mu\nu}$, i.e. $(M, \; g_{\mu\nu})$.

Let us consider a space-time where all null geodesics that start in a region $\cal{J}^{-}$ end at $\cal{J}^{+}$. Then, such a space-time, $(M,\; g_{\mu\nu})$, is said to contain a {\it black hole} if $M$ {\it is not} contained in the causal past\footnote{The notions of causal past and future of a space-time region can be found, for instance, in Hawking and Ellis (1973) and Wald (1984).}  $J^{-}({\cal{J^+}})$. In other words, there is a region from where no null geodesic can reach the {\it asymptotic flat}\footnote{Asymptotic flatness is a property of the geometry of space-time which means that in appropriate coordinates, the limit of the metric at infinity approaches the metric of the flat (Minkowskian) space-time.} future space-time, or, equivalently, there is a region of $M$ that is causally disconnected from the global future.  The {\it black hole region}, $BH$, of such space-time is $BH=[M-J^{-}({\cal{J^+}})]$, and the boundary of $BH$ in $M$, $H=J^{-}({\cal{J^+}}) \bigcap M$, is the {\it event horizon}. 

A black hole can be conceived as a space-time {\it region}, i.e. what characterizes the black hole is the metric and, consequently, space-time curvature. What is peculiar of this space-time region is that it is causally disconnected from the rest of the space-time: no events inside this region can make any influence on events outside. Hence the name of the boundary, event horizon: events inside the black hole are separated from events in the global external future of space-time. The events in the black hole, nonetheless, as all events, are causally determined by past events. A black hole does not represent a breakdown of classical causality.
     
The simplest type of black hole is described by the Schwarzschild metric; this metric characterizes the geometry of space-time outside a spherically symmetric matter distribution. The Schwarzschild metric for a static mass $M$ can be written in spherical coordinates $(t,r,\theta,\phi)$ as\footnote{These coordinates are usually referred as `Schwarzschild coordinates'.}:
\begin{equation}
	ds^{2}= \left(1-\frac{2GM}{rc^{2}}\right) c^{2}dt^2- \left(1-\frac{2GM}{rc^{2}}\right)^{-1} dr^{2} -r^{2} (d\theta^{2}+\sin^{2}\theta d\phi^{2}).\label{Schw}
\end{equation}

The radius
\begin{equation}
r_{\rm Schw}=\frac{2GM}{c^{2}},	\label{rS}
\end{equation}
is known as the {\sl Schwarzschild radius}. It corresponds to the event horizon if all the matter that generates the curvature is located at $r<r_{\rm Schw}$.

\begin{figure}
\centerline{\includegraphics[natwidth=1281,natheight=557,width=25pc]{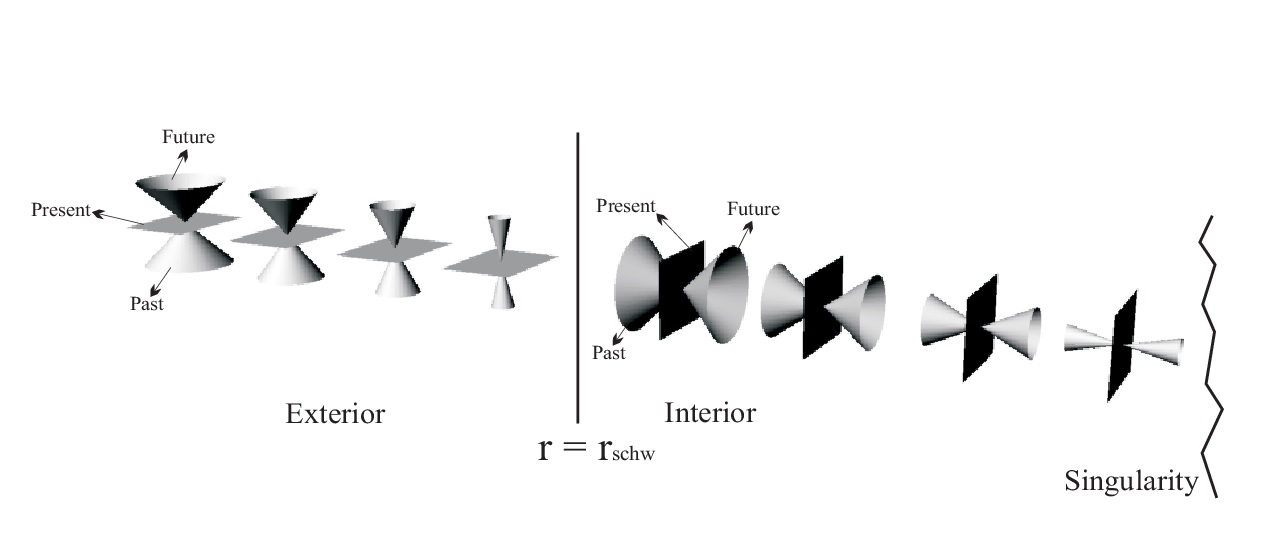}} 
\vspace{2pc}
\caption[]{Space-time diagram in Schwarzschild coordinates showing the light cones close to and inside a black hole.}
\label{Fig3}
\end{figure}

The light cones can be calculated from the metric (\ref{Schw}) imposing the null condition $ds^{2}=0$. Then: 
\begin{equation}
	\frac{dr}{dt}=\pm\left(1-\frac{2GM}{r}\right), \label{cones-Schw}
\end{equation}
where we made $c=1$. Notice that when $r\rightarrow \infty$, $dr/dt \rightarrow \pm 1$, as in Minkowski space-time. When $r\rightarrow 2GM$, $dr/dt \rightarrow 0$, and light moves along the surface $r=2GM$. The horizon is therefore a {\it null surface}. For $r<2GM$, the sign of the derivative is inverted. The inward region of $r=2GM$ is time-like for any physical system that has crossed the boundary surface. In  Fig. \ref{Fig3} we show the behavior of light cones in Schwarzschild coordinates. Similar figures are given in classical textbooks such as Misner et al. (1973, p. 848) and Carroll (2004, p. 219). As we approach to the horizon from the flat space-time region, the light cones become thinner and thinner indicating the restriction to the possible trajectories imposed by the increasing curvature. On the inner side of the horizon the local direction of time is `inverted' in the sense that null or time-like trajectories have in their future the singularity at the center of the black hole.


There is a very interesting consequence of all this: an observer on the horizon will have her present {\it along} the horizon. All events occurring on the horizon are simultaneous. The temporal distance from the observer at any point on the horizon to any event occurring on the horizon is zero (the observer is on a null surface $ds=0$ so the proper time interval is necessarily zero\footnote{Notice that this can never occur in Minkowski space-time, since there only photons can exist on a null surface. The black hole horizon, a null surface, can be crossed, conversely, by massive particles. The fact that the event horizon is a null surface is demonstrated in most textbook on relativity, see, e.g. Hartle (2003, p. 273) and d'Inverno (2002, p. 215).}). If the black hole has existed during the whole history of the universe, all events on the horizon during such history (for example the emission of photons on the horizon by infalling matter) are {\it present} to any observer crossing the horizon. These events are certainly not all present to an observer outside the black hole. If the outer observer is a presentist, she surely will think that some of these events do not exist because they occurred or will occur either in the remote past or the remote future. But if we accept that what there is cannot depend on the reference frame adopted for the description of the events, it seems we have an argument against presentism here. Before going further into the ontological implications, let us clarify a few physical points. 

What happens to an object when it crosses the event horizon? According to Eq. (\ref{Schw}), there is a singularity at $r=r_{\rm Schw}$. However, the metric coefficients can be made regular by a change of coordinates. For instance we can consider Eddington-Finkelstein coordinates. Let us define a new radial coordinate $r_{*}$ such that radial null rays satisfy $d(c t\pm r_{*})=0$. Using Eq. (\ref{Schw}) it can be shown that:
$$r_{*}=r +  \frac{2GM}{c^{2}} \log \left|\frac{r-2GM/c^{2}}{2GM/c^{2}}\right|.$$
Then, we introduce: 
$$v=ct+r_{*}.$$ The new coordinate $v$ can be used as a time coordinate replacing $t$ in Eq. (\ref{Schw}). This yields: $$ ds^{2}=\left(1-\frac{2GM}{rc^{2}}\right) (c^2dt^{2}-dr^{2}_{*})-r^{2} d\Omega^{2},$$ or
\begin{equation}
ds^{2}=\left(1-\frac{2GM}{rc^{2}}\right) dv^{2}-2drdv -r^{2} d\Omega^{2}, \label{EF}
\end{equation}
where $$d\Omega^{2}=d\theta^{2}+\sin^{2} \theta d\phi^{2}.$$



\begin{figure}
\centerline{\includegraphics[natwidth=423,natheight=657,width=15pc]{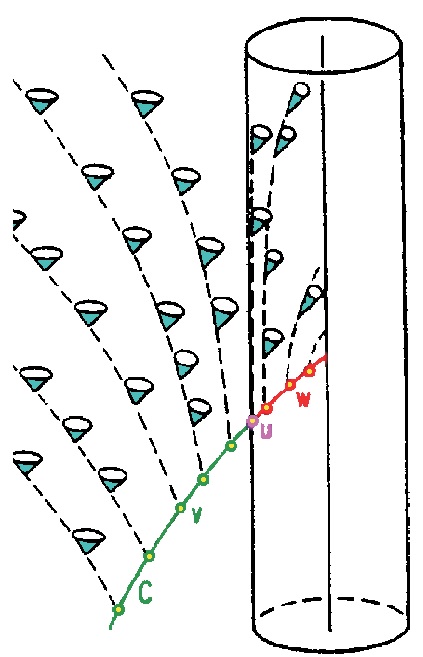}} 
\vspace{2pc}
\caption[]{Space-time diagram in Eddington-Finkelstein coordinates showing the light cones close to and inside a black hole. The event $u$ occurs at the event horizon, represented by the surface of the cylinder. The global time direction is in the direction of the axis of the cylinder.}
\label{Fig4}
\end{figure}

Notice that in Eq. (\ref{EF}) the metric is non-singular at $r=2GM/c^{2}$. The only real singularity is at $r=0$, since the Riemann tensor diverges there. In order to plot the space-time in a $(t,\; r)$-plane, we can introduce a new time coordinate $t_{*}=v-r$. From the metric (\ref{EF}) or from Fig. \ref{Fig4} we see that the line $r=r_{\rm Schw}$, $\theta=$constant, and $\phi=$ constant is a null ray, and hence, the surface at $r=r_{\rm Schw}$ is a null surface. This null surface is an event horizon because inside $r=r_{\rm Schw}$ all cones have $r=0$ in their future. In Figure \ref{Fig4} we see how the light cones tilt in Eddington-Finkelstein coordinates. {\sl All} light cones of events at $r=r_{\rm Schw}$ have their null surfaces coincident with the horizon. The horizon itself is a null surface common to all light cones associated with physical systems that are entering the black hole.

We remark, then, that the horizon 1) does not depend on the choice of the coordinate system adopted to describe the black hole, 2) it is an absolute null surface, in the sense that this property is intrinsic and not frame-dependent, and 3) is non-singular (or `well-behaved', i.e. space-time is regular on the horizon).

More complex black holes can be considered. For instance, rotating (Kerr), charged (Reissner-Nordstr\"om), or both rotating and charged (Kerr-Newman) black holes. In all of them the outer event horizons are null surfaces, so the above considerations remain valid. Rotating black holes introduce additional problems to presentism, related to the existence of Cauchy horizons and closed timelike curves, but we do not discuss these issues here. Rather, we focus on the implications of the existence of null surfaces that can be crossed by massive particles or observers for presentism.  


\section{Ontological implications}

In a world described by special relativity, the only way to cross a null surface is by moving faster than the speed of light. As we have seen, this is not the case in a universe with black holes. We can then argue against presentism along the following lines. \\ 

Argument $A1$:

\begin{itemize}
	\item{$P1$: There are black holes in the universe.}
	\item{$P2$: Black holes are correctly described by general relativity.}
	\item{$P3$: Black holes have closed null surfaces (horizons).}
	
	\item{Therefore, there are closed null surfaces in the universe.}
\end{itemize}

Argument $A2$:

\begin{itemize}
	\item{$P4$: All events on a closed null surface are simultaneous with any event on the same surface.}
	\item{$P4i$: All events on the closed null surface are simultaneous with the birth of the black hole.}
	\item{$P5$: Some distant events are simultaneous with the birth of the black hole, but not with other events related to the black hole.}
	
	\item{Therefore, there are events that are simultaneous in one reference frame, and not in another.}
\end{itemize}

Simultaneity is frame-dependent. Since what there exist cannot depend on the reference frame, we conclude that there are non-simultaneous events. Therefore, presentism is false.

Let us see which assumptions are open to criticism by the presentist. 

An irreducible presentist might plainly reject $P1$. Although there is significant astronomical evidence supporting the existence of black holes (e.g. Casares 2006, Camenzind 2007, Paredes 2009, Romero and Vila 2013), the very elusive nature of these objects still leaves room for some speculations like gravastars, and other exotic compact objects. The price of rejecting $P1$, however, is very high: black holes are now a basic component of most mechanisms that explain extreme events in astrophysics, from quasars to the so-called gamma-ray bursts, from the formation of galaxies to the production of jets in binary systems. The presentist rejecting black holes should reformulate the bulk of contemporary high-energy astrophysics in terms of new mechanisms. In any case, $P1$ is susceptible of empirical validation through direct imagining of the super-massive black hole ``shadow'' in the center of our galaxy by sub-mm interferometric techniques in the next decade (e.g. Falcke et al. 2011). In the meanwhile, the cumulative case for the existence of black holes is overwhelming, and very few scientists would reject them on the basis of metaphysical considerations only.

The presentist might, instead, reject $P2$. After all, we {\it know} that general relativity fails at the Planck scale. Why should it provide a correct description of black holes? The reason is that the horizon of a black hole is quite far from the region where the theory fails (the singularity). The distance, in the case of a Schwarzschild black hole, is $r_{\rm Schw}$ (see Eq. \ref{rS}). For a black hole of 10 solar masses, as the one suspected to form part of the binary system Cygnus X-1, this means $30$ km. And for the black hole in the center of the galaxy, the Schwarzschild radius is of about 12 million km. Any theory of gravitation must yield the same results as general relativity at such distances. So, even if general relativity is not the right theory for the classical gravitational field, the correct theory should predict the formation of black holes under the same conditions.



\begin{figure}
\centerline{\includegraphics[natwidth=1341,natheight=485,width=30pc]{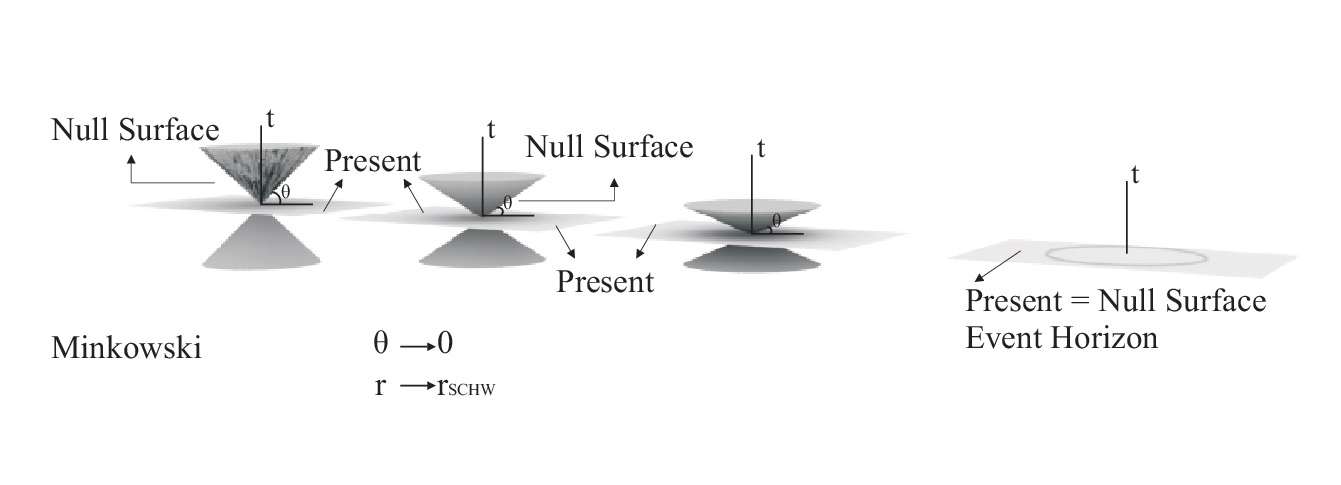}} 
\vspace{2pc}
\caption{Light cones aperture angles at different distances from the horizon of a Schwarzschild black hole. On the horizon the null surface is coincident with the hyperplane of present.} \label{Fig5}
\end{figure}

There is not much to do with $P4$, since it follows from the condition that defines the null surface: $ds=0$\footnote{ $ds=cd\tau=0 \rightarrow d\tau=0$, where $d\tau$ is the proper temporal separation.}; similarly $P4i$ only specifies one of the events on the null surface. A presentist might refuse to identify `the present' with a null surface. After all, in Minksowskian space-time or even in a globally time-orientable pseudo-Riemannian space-time the present is usually taken as the hyperplane perpendicular to the local time (see Figs. \ref{Fig1} and \ref{Fig2}). But in space-times with black holes, the horizon is not only a null surface, it is also a surface locally normal to the time direction. This can be appreciated in Figure \ref{Fig5}, where the angle $\theta$ is the angle between the null surface and the hyperplane of the present. In a Minkowskian space-time such an angle is 45$^\circ$ when the speed of light is measured in natural units ($c=1$). In such a space-time, certainly the plane of the present is not coincident with a null surface. However, close to the event horizon of a black hole, things change, as indicated by Eq. (\ref{cones-Schw}). As we approach the horizon, $\theta$ goes to zero, i.e. the null surface matches the plane of the present. On the horizon, both surfaces are exactly coincident: $\theta\rightarrow 0$ when $r\rightarrow r_{\rm Schw}$. A presentist rejecting the identification of the present with a {\it closed} null surface on an event horizon should abandon what is perhaps her most cherished belief: the identification of `the present' with hypersurfaces that are normal to a local time-like direction.

The result mentioned above is not a consequence of any particular choice of coordinates but, as mentioned in Sect. \ref{BH}, an intrinsic property of a black hole horizon. This statement can be easily proved. The symmetries of Schwarzschild space-time imply the existence of a preferred radial function, $r$, which serves as an affine parameter along both null directions. The gradient of this function, $r_{a}=\nabla_{a} r$ satisfies ($c=G=1$):

\begin{equation}
r^{a}r_{a}=\left(1-\frac{2M}{r}\right). \label{ra}
\end{equation}
Thus, $r^{a}$ is space-like for $r>2M$, null for $r=2M$, and time-like for $r<2M$. The 3-surface given by $r=2M$ is the horizon $H$ of the black hole in Schwarzschild space-time. From Eq. (\ref{ra}) it follows that $r^{a}r_{a}=0$ over $H$, and hence $H$ is a null surface\footnote{An interesting case is Schwarzschild space-time in the so-called Painlev\'e-Gullstrand coordinates. In these coordinates the interval reads: 
\begin{equation}
	ds^{2}=dT^{2}-\left(dr + \sqrt{\frac{2M}{r}} dT\right)^{2} - r^{2}d\Omega^{2},\label{PG}
\end{equation}
with
\begin{equation}
	T=t + 4M \left(\sqrt{\frac{2M}{r}} + \frac{1}{2} \ln \left| \frac{\sqrt{\frac{2M}{r}}-1}{\sqrt{\frac{2M}{r}}+1} \right|\right).
\end{equation}

If a presentist makes the choice of identifying the present with the surfaces of $T=$constant, from Eq. (\ref{PG}): $ds^{2}= - dr^{2} - r^{2}d\Omega^{2}$. Notice that for $r=2M$ this is the event horizon, which in turn, is a null surface. Hence, with such a choice, the presentist is considering that the event horizon is the hypersurface of the present, for all values of $T$. This choice of coordinates makes particularly clear that the usual presentist approach to define the present in general relativity self-defeats her position if space-time allows for black holes.}.  

Premise $P5$, perhaps, looks more promising for a last line of presentist defense. It might be argued that events on the horizon are not simultaneous with any event in the external universe. They are, in a very precise sense, cut off from the universe, and hence cannot be simultaneous with any distant event. Let us work out a counterexample. 

\begin{figure}
\centerline{\includegraphics[natwidth=1055,natheight=540,width=25pc]{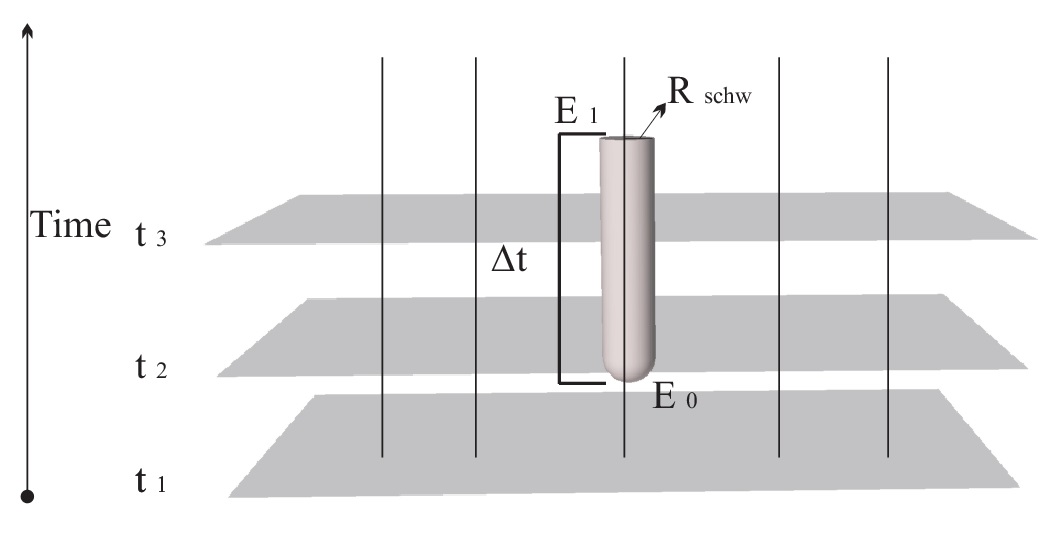}} 
\vspace{2pc}
\caption[]{Foliation of a temporally-oriented space-time with a black hole. The time span from the birth of the black hole to the event $E_{1}$ is $\normalsize\Delta t$.}
\label{Fig6}
\end{figure}

The so-called long gamma-ray bursts are thought to be the result of the implosion of a very massive and rapidly rotating star. The core of the star becomes a black hole, which accretes material from the remaining stellar crust. This produces a growth of the black hole mass and the ejection of matter from the magnetized central region in the form of relativistic jets (e.g. Woosley 1993). Approximately, one of these events occur in the universe per day. They are detected by satellites such as {\it Swift} (e.g. Piran and Fan 2007), with durations of a few tens of seconds. This is the time that takes for the black hole to swallow the collapsing star. Let us consider a gamma-ray burst of, say, 10 seconds. Before these 10 seconds, the black hole did not exist for a distant observer $O1$. Afterwards, there is a black hole in the universe that will last more than the life span of any human observer. Let us now consider an observer $O2$ collapsing with the star. At some instant she will cross the null surface of the horizon. This will occur within the 10 seconds that the collapse lasts for $O1$. But for $O2$ all photons that cross the horizon are simultaneous, including those that left $O1$ long after the 10 seconds of the event and crossed the horizon after traveling a long way. For instance, photons  leaving the planet of $O1$ one million years after the gamma-ray burst, might cross the horizon, and then can interact with $O2$. So, the formation of the black hole is simultaneous with events in $O1$ and $O2$, but these very same events of $O2$ are simultaneous with events that are in the distant future of $O1$. The situation is illustrated in Fig. \ref{Fig6}.


The reader used to work with Schwarzschild coordinates perhaps will object that $O2$ never reaches the horizon, since the approaching process takes an infinite time in a distant reference frame, like that of $O1$. This is, however, an effect of the choice of the coordinate system and the test-particle approximation (see, for instance, Hoyng 2006, p.116). If the process is represented in Eddington-Finkelstein coordinates, it takes a finite time for the whole star to disappear, as shown by the fact that the gamma-ray burst are quite short events. Accretion/ejection processes, well-documented in active galactic nuclei and microquasars (e.g. Mirabel et al. 1998) also show that the time taken to reach the horizon is finite in the asymptotically flat region of space-time.   

Our conclusion is that presentism, as usually stated, provides a defective picture of the ontological substratum of the world.

\section{Final remarks}

What kind of ontological view is compatible with black hole astrophysics? We suggest that one where what we call `present' has a local rather than a global character. The detachment between local and global time occurring in a black hole results, as we have seen, in a checkmate for the presentist position. There is no way to re-identify the present to give it back a global meaning in a universe with black holes. 

The intuitive ontology adopted by most practising astrophysicists is one where there are things, and these things change relative to each other. One can speculate that space-time is an emergent property of the system of all things (e.g. Bunge 1977, Perez-Bergliaffa et al. 1998). The exact formulation of such an ontological theory to encompass a relativistic view of the world, taking into account the peculiarities of non-local effects in quantum mechanics, is an open problem. A problem that cries for attention from both philosophers and scientists alike.

%

\begin{thebibliography}{}

\bibitem{Bigelow-1996-Tomberling}
Bigelow, J. 1996. Presentism and Properties. In Metaphysics, Vol.  10  of Philosophical Perspectives, ed. J.E. Tomberlin, 35�52. Cambridge, Mass.:
Blackwell.


\bibitem{Bunge-77-Dordrecht}
Bunge, M. 1977. Treatise of Basic Philosophy. Ontology I: The Furniture of the World. Dordrecht: Reidel.

\bibitem{Came-07-Springer}
Camenzind, M. 2007. Compact objects in Astrophysics : White Dwarfs, Neutron Stars and Black Holes. Berlin: Springer-Verlag.

\bibitem{Carroll-2004}
Carroll, S. 2004. Spacetime and Geometry. An Introduction to General Relativity. San Francisco: Addison Wesley.

\bibitem{Casa-05-MNRAS}
Casares, J. 2006. Observational Evidence for Stellar Mass Black Holes. In Proceedings IAU Symposium No 238, eds. V. Karas $\&$ G. Matt, Volume 2, 3-12.


\bibitem{Chil-90-Nous}
Chisholm, R. 1990. Events Without Time: An Essay On Ontology. No$\hat{u}$s 24: 413-428. 



\bibitem{Craig-01-Dordrecht}
Craig, W.L. 2001. Time and the Metaphysics of Relativity. Dordrecht: Kluwer.



\bibitem{Crisp-03-Oxford}
Crisp, T. 2003.
Presentism. In The Oxford Handbook of Methaphysics, eds. M. J. Loux \& D. W. Zimmerman, 211-245. Oxford: Oxford University Press.

\bibitem{Crisp-07-in}
Crisp, T. 2007.
Presentism, Eternalism and Relativity Physics. In Einstein, Relativity and Absolute Simultaneity, eds. W. L. Craig \& Q. Smith, 262-278. London: Routledge. 

\bibitem{DInverno-1992}
d'Inverno, R. 2002. Introducing Einstein's Relativity. Oxford: Clarendon Press.

\bibitem{Dolev-2006}
Dolev, Y. 2006. How to Square a Non-localized Present with Special Relativity. In The Ontology of Spacetime, ed. D. Dieks, 177-190. The Netherlands: Elsevier.

\bibitem{Dorato-2006}  
Dorato M. 2006. The Irrelevance of the Presentist/Eternalist Debate for the Ontology of Minkowski Spacetime. In The Ontology of Spacetime, ed. D. Dieks, 93-109. The Netherlands: Elsevier.

\bibitem{Einstein-SR}
Einstein, A. 1905. Zur Elektrodynamik bewegter K$\ddot{o}$rper. Annalen der Physik 17: 891-921.

\bibitem{Einstein-PAW}
Einstein, A. 1907. Relativit$\ddot{a}$tsprinzip und die aus demselben gezogenen Folgerungen. Jahrbuch der Radioaktivit$\ddot{a}$t 4: 411-462.


\bibitem{Einstein-PAW}
Einstein, A. 1915. Die Feldgleichungen der Gravitation. Preussische Akademie der Wissenschaften, 844-847. 

\bibitem{Ellis-2006}
Ellis, G.F.R. 2006. Physics in the Real Universe: Time and Spacetime. General Relativity and Gravitation, 38: 1797-1824.

\bibitem{Ellis-2010}
Ellis, G.F.R., Rothman, T. 2010. Time and Spacetime: The Crystallizing Block Universe. International Journal of Theoretical Physics, 49: 988-1003.


\bibitem{Gru-73-Rei}
Gr$\ddot{u}$nbaum, A. 1973. Philosophical Problems of Space and Time. Dordrecht: Reidel.

\bibitem{Falcke-2011-IAU}
Falcke, H., Markoff S., Bower G. C., Gammie, C. F., Moscibrodzka, M. \& Maitra, D. 2011. The Jet in the Galactic Center: An Ideal Laboratory for Magnetohydrodynamics and General Relativity. In Jets at all Scales, Proceedings of the International Astronomical Union, IAU Symposium, eds. G. E. Romero, R. A. Sunyaev $\&$ T. Belloni, Volume 275, 68-76.

\bibitem{Harrington-2008}
Harrington, J. 2008. Special Relativity and The Future: A Defense of the Point-Present. Studies in the History and Philosophy of Modern Physics, 39: 82-101.

\bibitem{Harrington-2009}
Harrington, J. 2009. What ``Becomes'' in Temporal Becoming? American Philosophical Quarterly, 46: 249-265.

\bibitem{Hartle-2002}
Hartle, J. B. 2003. Gravity: An Introduction to Einstein's General Relativity. San Francisco: Addison Wesley.


\bibitem{Hawking-73-Cambridge}
Hawking, S. \& Ellis, G.F.R. 1973.
The Large-Scale Structure of Space-Time. Cambridge: Cambridge University Press.

\bibitem{Hi-2000-Proc}
Hinchliff, M. 2000.
A Defense of Presentism in a Relativistic Setting. Philosophy of Science (Proceedings) LXVII, S575-S586.

\bibitem{Hoyng-2006-PP}
Hoyng, S. 2006. Relativistic Astrophysics and Cosmology: A Primer. Berlin: Springer.


\bibitem{Ma-04-Zi}
Markosian, N. 2004.
A Defense of Presentism. In Oxford Studies in Methaphysics, ed. D. Zimmerman, 1: 47-82. Oxford: Oxford University Press. 


\bibitem{Maxwell-1985}
Maxwell, N. 1985. Are Probabilism and Special Relativity Imcompatible? Philosophy of Science, 52: 23-43.


\bibitem{Me-99-Nous}
Merricks, T. 1999.
Persistance, Parts and Presentism. No$\hat{u}$s 33: 421-438. 


\bibitem{Mink-1900-al}
Minkowski, H. 1907.
Die Grundgleichungen f$\ddot{u}$r die elektromagnetischen Vorg$\ddot{a}$nge in bewegten K$\ddot{o}$rpern, Nachrichten von der Gesellschaft der Wissenschaften zu G$\ddot{o}$ttingen, Mathematisch-Physikalische Klasse, 53�111. 




\bibitem{Mink-1900-al}
Minkowski, H. 1909.
Lecture ``Raum und Zeit, 80th Versammlung Deutscher Naturforscher (K$\ddot{o}$ln, 1908)''. Physikalische Zeitschrift 10: 75-88. 

\bibitem{Mira-1998-al} 
Mirabel, I.F., Dhawan, V., Chaty, S., Rodriguez, L. F., Marti, J.,Robinson, C. R.,Swank, J., Geballe, T. 1998.
 Accretion instabilities and jet formation in GRS 1915+105. Astronomy  $\&$ Astrophysics 330: L9-L12.
 
\bibitem{MTW-1973}
Misner, C.W., Thorne, K.S., Wheeler, J.A. 1973. Gravitation. W.H. Freeman. 

\bibitem{Zi-2011-PP} 
Mozersky, M. J. 2011. Presentism. In The Oxford Handbook of Philosophy of Time, ed. C. Callender, 122-144. Oxford: Oxford University Press.

\bibitem{Norton-2013}
Norton, J. D. 2013. The Burning Fuse Model of Unbecoming in Time. Paper presented at the Workshop on Cosmology and Time. Penn State University.


\bibitem{Paredes-2009} 
Paredes, J. M. 2009. Black Holes in the Galaxy. In Compact Objects and their Emission, Argentinian Astronomical Society Book Series, eds. G. E. Romero $\&$ P. Benaglia, Volume 1, 91-121. 

\bibitem{Piran-2007-al} 
Piran, T. \& Fan, Y. 2007. Gamma-Ray Burst Theory after Swift. Philosophical Transactions of the Royal Society A 365: 1151-1162.


\bibitem{Plebanski-06-Cambridge}
Pleba\'nski, J. \& Krasi\'nski, A. 2006. An Introduction to General Relativity and Cosmology. Cambridge: Cambridge University Press.

\bibitem{Poincare-02-Paris}
Poincar\'e, H. 1902. La Science et l'Hypoth\`ese. Paris: E. Flammarion.







\bibitem{PerezRomeroVucetich} Perez-Bergliaffa, S.E, Romero, G.E. \& Vucetich, H. 1998. Toward an axiomatic pregeometry of space-time. International Journal of Theoretical Physics 37: 2281-2298.

\bibitem{Put-1967-JP}
Putnam, H. 1967. Time and Physical Geometry. Journal of Philosophy 64: 240-247. 

\bibitem{Rea-03-Oxford}
Rea, M. C. 2003. Four-Dimensionalism. In The Oxford Handbook of Methaphysics, eds. M. J. Loux \& D. W. Zimmerman, 246-80. Oxford: Oxford University Press.

\bibitem{Rie-1966-PP} 
Rietdijk, C. W. 1966. A Rigorous Proof of Determinism Derived from this Special Theory of Relativity. Philosophical Papers 33: 341-344. 

\bibitem{Romero-Vila-2013}
Romero, G.E., Vila, G.S. 2013. Introduction to Black Hole Astrophysics. Lectures Notes in Physics. Berlin: Springer.

\bibitem{Savitt-2006}
Savitt, S. S. 2006. Presentism and Eternalism in Perspective. In The Ontology of Spacetime, ed. D. Dieks, 111-127. The Netherlands: Elsevier.

\bibitem{Sau-02-Ca}
Saunders, S. 2002. How Relativity Contradicts Presentism. In Time, Reality $\&$ Experience, Royal Institute of Philosophy, Supplement, ed. C. Callender, 277-292. Cambridge, New York: Cambridge University Press. 

\bibitem{Smart-63-Ca}
Smart, J. J. C. 1963. Philosophy and Scientific Realism. London: Routledge.

\bibitem{Stein-1968-JP}
Stein, H. 1968. On Einstein-Minkowski Space-Time. Journal of Philosophy 65: 5-23. 

\bibitem{Stein-1991-PP}
Stein, H. 1991. On Relativity Theory and Openness of the Future. Philosophy of Science 58: 147-167. 

\bibitem{Wald-1984-PP} 
Wald, R.M. 1984. General Relativity. Chicago: The University of Chicago Press. 



\bibitem{Woosley-1993-PP}
Woosley, S. E. 1993. Gamma-Ray Bursts from Stellar Collapse to a Black Hole?. Bulletin of the American Astronomical Society 25, 894.


\bibitem{Zi-1996-PP}
Zimmerman, D. 1996. Persistence and Presentism. Philosophical Papers 1996: 35-52. 
 

\bibitem{Zi-2011-PP}
Zimmerman, D. 2011. Presentism and the Space-Time Manifold. In The Oxford Handbook of Philosophy of Time, ed. C. Callender, 163-244. Oxford: Oxford University Press.




\end{thebibliography}

%
%

\end{document}